\documentclass[journal=jacsat,manuscript=article]{achemso}

\usepackage[version=3]{mhchem} 

\author{Marius Schulte}
\affiliation[Ludwig-Maximilians-Universit{\"a}t
M{\"u}nchen]{Ludwig-Maximilians-Universit{\"a}t M{\"u}nchen, Department Chemie,
Butenandtstr. 11, 81377 M{\"u}nchen, Germany}
\author{Irmgard Frank}
\email{irmgard.frank(at)theochem.uni-hannover.de}
\affiliation[Leibniz Universit{\"a}t Hannover]{Leibniz Universit{\"a}t
Hannover, Institut f{\"u}r Physikalische Chemie und Elektrochemie, Callinstr.
3A, 30167 Hannover, Germany}

\title{Car-Parrinello Simulation of the Reaction of Aluminium with Oxygen}

\begin{document}
\begin{abstract}
We present Car-Parrinello molecular dynamics simulations of the initial reaction steps leading to an inert oxide layer on aluminium. The mechanism of the reaction of the aluminium surface with single oxygen molecules is analysed. After adsorption at the surface the oxygen molecules dissociate at a femtosecond timescale and the atoms are chemisorbed at the surface at a distance of several angstrom. When the aluminium surface is exposed to higher oxygen pressure, a surface layer essentially consisting of threefold coordinated oxygen atoms starts to form.
\end{abstract}

\pagebreak
\section{Introduction}

Aluminium 
as a base metal gains its corrosion resistance from a thin layer consisting of a mixture of aluminium oxide and aluminium
hydroxide which forms immediately when a clean aluminium surface is exposed to air \cite{Holleman1995}. The
composition, thickness and properties of this chemically passivating layer
depend on the particular conditions of formation.  Spontaneous self-passivation
leads to layers of a few nanometers thickness. Significantly thicker and harder
layers essentially consisting of well-ordered $\alpha$-Al$_2$O$_3$ can be obtained electrochemically.
Like for bulk aluminium oxide, the surface of passivated aluminium is
covered with OH groups with different coordination to the bulk
\cite{Knoezinger1978}. The reaction of oxygen with aluminium was investigated in detail with
STM experiments as early as 1992 emphasizing the role of 'hot adatoms'
\cite{Brune1992,Brune1993}. The oxygen atoms were found to be at an
average distance of 8 nm after dissociation. Theoretical studies
yielded lower separations \cite{Engdahl1994,Wahnstrom1996}.
A subsequent experimental study \cite{Schmid2001}
measured a much lower transient mobility of the adsorbed oxygen atoms (oxygen-oxygen separation of 0.5 nm on average)
which is in way better agreement with the high oxygen affinity of aluminium.
The subject continued to be investigated in experimental and theoretical studies. 
A recent HRTEM study investigates the growth of an aluminium oxide
layer in contact with the melt \cite{Oh2010}.

In the present study we want to simulate, at first-principles level, the
initial steps of the formation of such layers.  Previous theoretical
investigations using density functional theory (DFT) focussed on alumina surfaces and
their reactivity
\cite{Causa1989,Manassidis1993,Frank1995,Frank1996,Hass1998,Hass2000}.  Early
work confirmed the experimental finding that the (0001) surface is most stable
\cite{Causa1989,Manassidis1993}. First-principles molecular dynamics
simulations \cite{Hass1998,Hass2000} showed the facile reaction of the oxide
surface with water molecules leading to OH coverage. In a study of the reaction of
oxygen with aluminium \cite{Behler2005,Behler2008} the authors induced an artificial barrier
to chemisorption in order to explain the results by Brune et al. \cite{Brune1992}.
This barrier was constructed by claiming the relevance of 'non-adiabatic effects'
when the triplet oxygen molecule chemisorbs at the surface. However, the proper treatment of
non-adiabatic effects with density functional theory in the Kohn-Sham approximation is unclear at best.
On top, it is not at all clear how any non-adiabatic calculation should help to describe
an intersystem crossing as it does not include spin terms in the Hamiltonian.
Also in view of the more recent experiments which indicate a much lower transient mobility and faster
chemisorption \cite{Schmid2001},
an artificial extension of DFT might not be necessary to model the system. 
To analyse the mechanism and to elucidate the influence of the multiplicity we investigate the reaction of an aluminium surface with
one oxygen molecule simulating the gas-phase situation as well as the reaction with liquid oxygen.


\section{Results and discussion}

\begin{figure}
  \includegraphics[width=8cm]{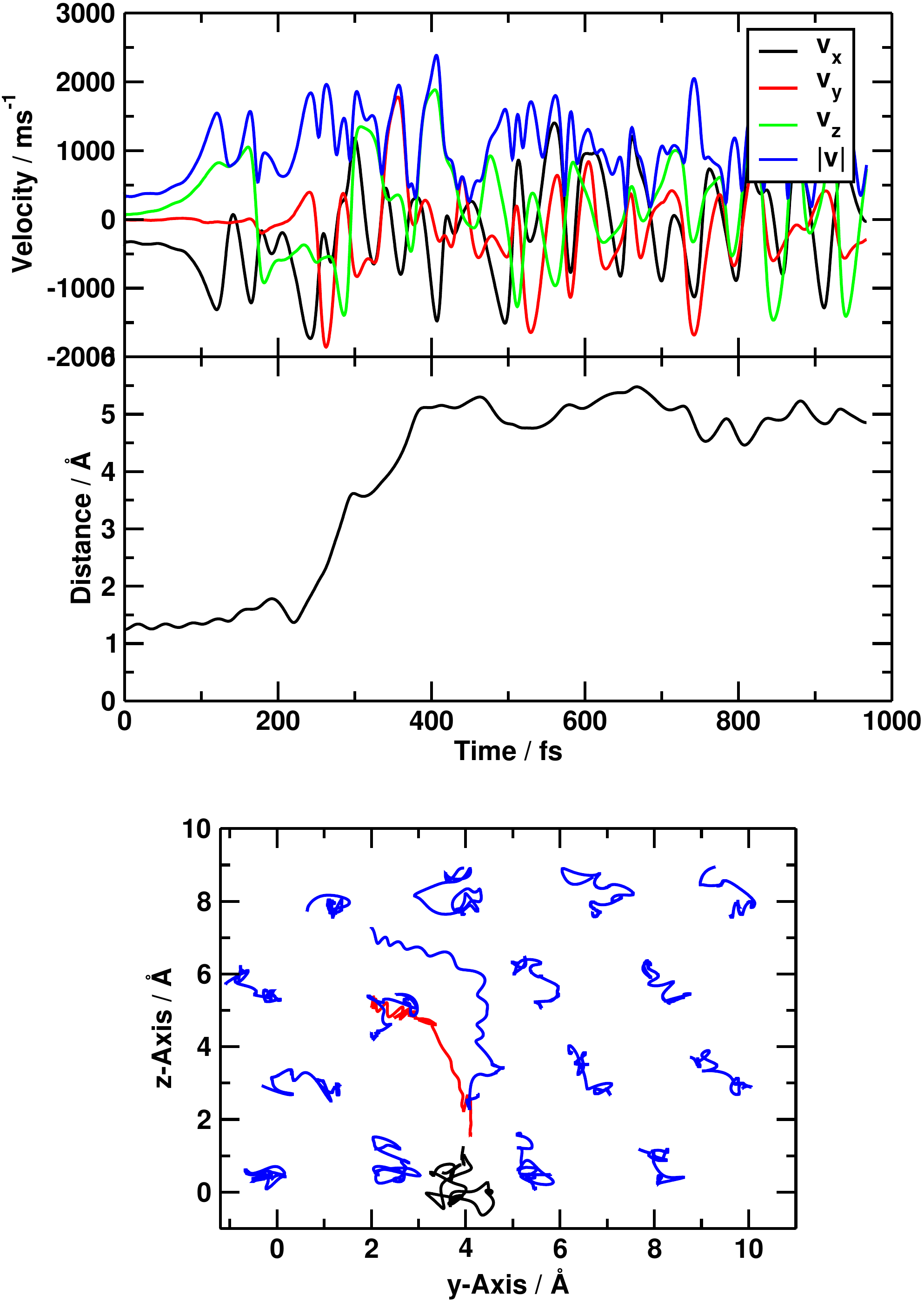}
  \caption{Reaction of an oxygen molecule with an aluminium surface as observed in one of the simulation runs.
The upper plot shows the velocities of the center of mass of the two oxygen
atoms and their distance. In the lower plot the motion of the oxygen atoms
on the aluminium surface are depicted (black and red). One of the surface atoms (blue)
is moved to another lattice site as a consequence of the impact.
}
  \label{Fig1} 
\end{figure}

A series of simulations with one attacking oxygen only was performed at a temperature of 300 K
using different initial orientations of the oxygen molecule relative to the surface.
The incident oxygen molecule is moving towards the surface
with a velocity of 400 m/s which leads to a reaction within a few hundred femtoseconds in all ten simulation runs. 
In each case, the oxygen molecule binds to the surface and dissociates.
\ref{Fig1} illustrates the motion of the two oxygen atoms for one of the simulation runs.
The molecule hits the surface 120 fs after the end of the equilibration. At this point
it is accelerated to a velocity of roughly 1300 m/s by the attraction of the surface.
The distance plot shows
that it starts to dissociate but 100 fs later.
Soon after the first oxygen atom (red graph in \ref{Fig1}, lower plot) contacts the surface,
the second atom (black graph) is bound to a second
surface aluminium atom.
In this particular simulation run the aluminium layer is strongly disturbed
immediately upon the oxygen impact:
During the bond dissociation the first atom is pushed over the bound
aluminium
atom to its new position while the aluminium atom itself is pulled out of
the
surface to another position leading to a relatively large aluminium displacement (\ref{Fig1}).
Apart from this special feature of this particular MD run, the reaction follows always the
same scheme: adsorption of one oxygen atom, adsorption of the second, dissociation and
relaxation at distant lattice sites.

\begin{figure}
  \includegraphics[width=8cm]{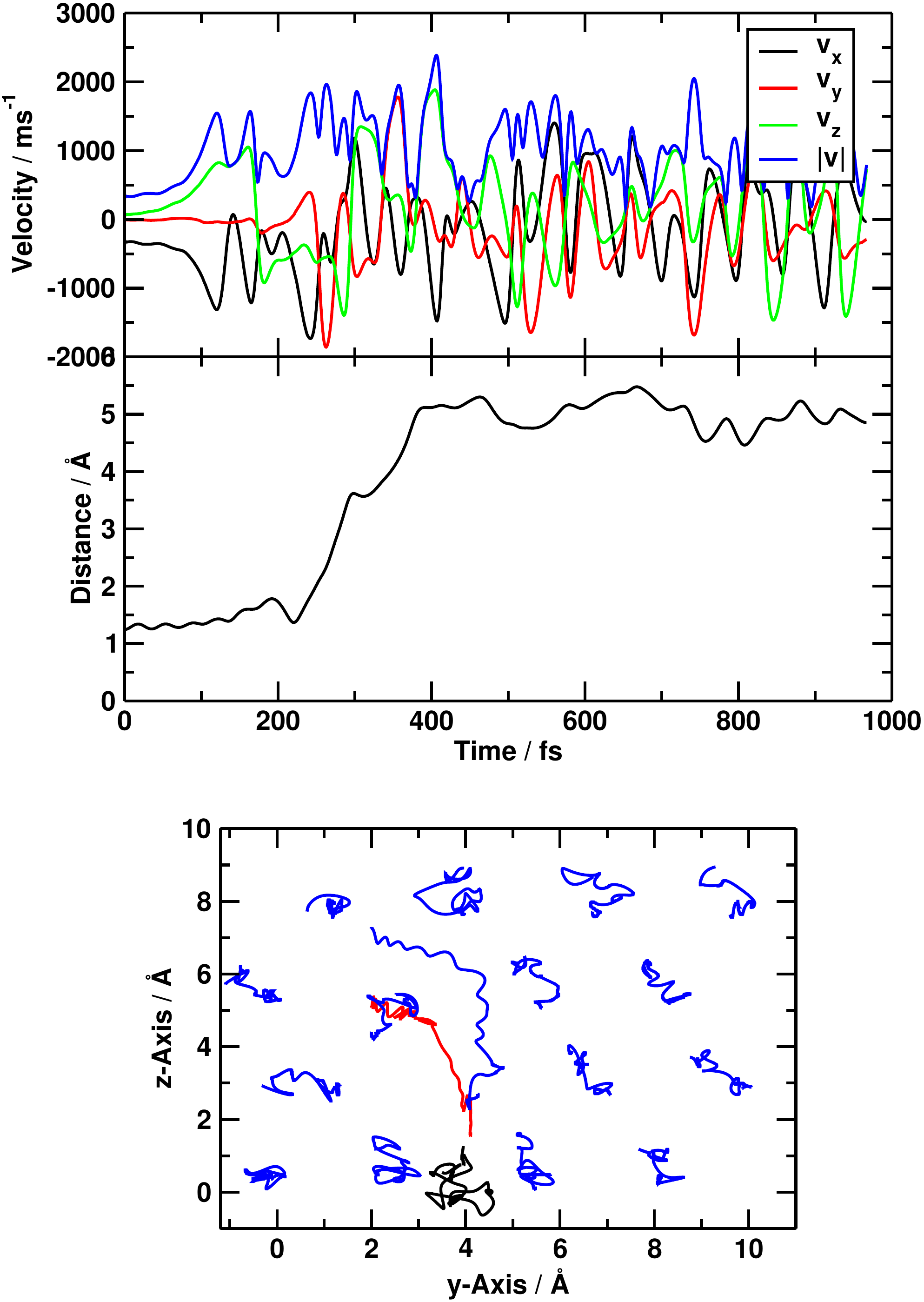}
  \caption{Change of energy, temperature, charge and spin charge during the strongly
exothermic reaction of a
single oxygen molecule with the surface. Top graph: Kohn-Sham energy (black),
classical energy (red), total energy of the Car-Parrinello Lagrangian (green),
and temperature (blue). The temperature is proportional to the kinetic energy of the ions
and hence to the difference between the black and the red curve, while the difference between
the red and the green curve corresponds to the fictitious kinetic energy of the orbitals
in Car-Parrinello theory. The two lower graphs show the charges and spin charges
of the two oxygen atoms. While the spin charge at the oxygen
atoms decreases during adsorption and is transferred to the surface, the
charge of the oxygen atoms is increased during the full reaction including dissociation
and relaxation.
}
  \label{Fig2}
\end{figure}

\ref{Fig2} shows the increase in temperature (blue graph) of the total system. Both the adsorption
and the consecutive dissociative reaction lead to a significant increase
in temperature. From the comparison with \ref{Fig1} it is obvious that initially
the oxygen atoms gain kinetic energy, while in the further course of the reaction
the increase of the kinetic energy is taken up by the surface.
The Kohn-Sham energy is lowered by roughly 0.14 a.u. (370 kJ/mol). Since no
thermostats were used in the simulation, the electronic system
heats up quite a bit and gains kinetic energy. 
The single reaction steps can be followed from the graphs of the charge and the
spin charge (\ref{Fig2}).
During the first reaction step, which is the adsorption to the surface, the charge of
the oxygen molecule changes only partially. The spin charge is transferred
to the surface within 160 fs while the oxygen atoms are fully ionized after about
400 fs. 

\begin{figure}
  \includegraphics[width=8cm]{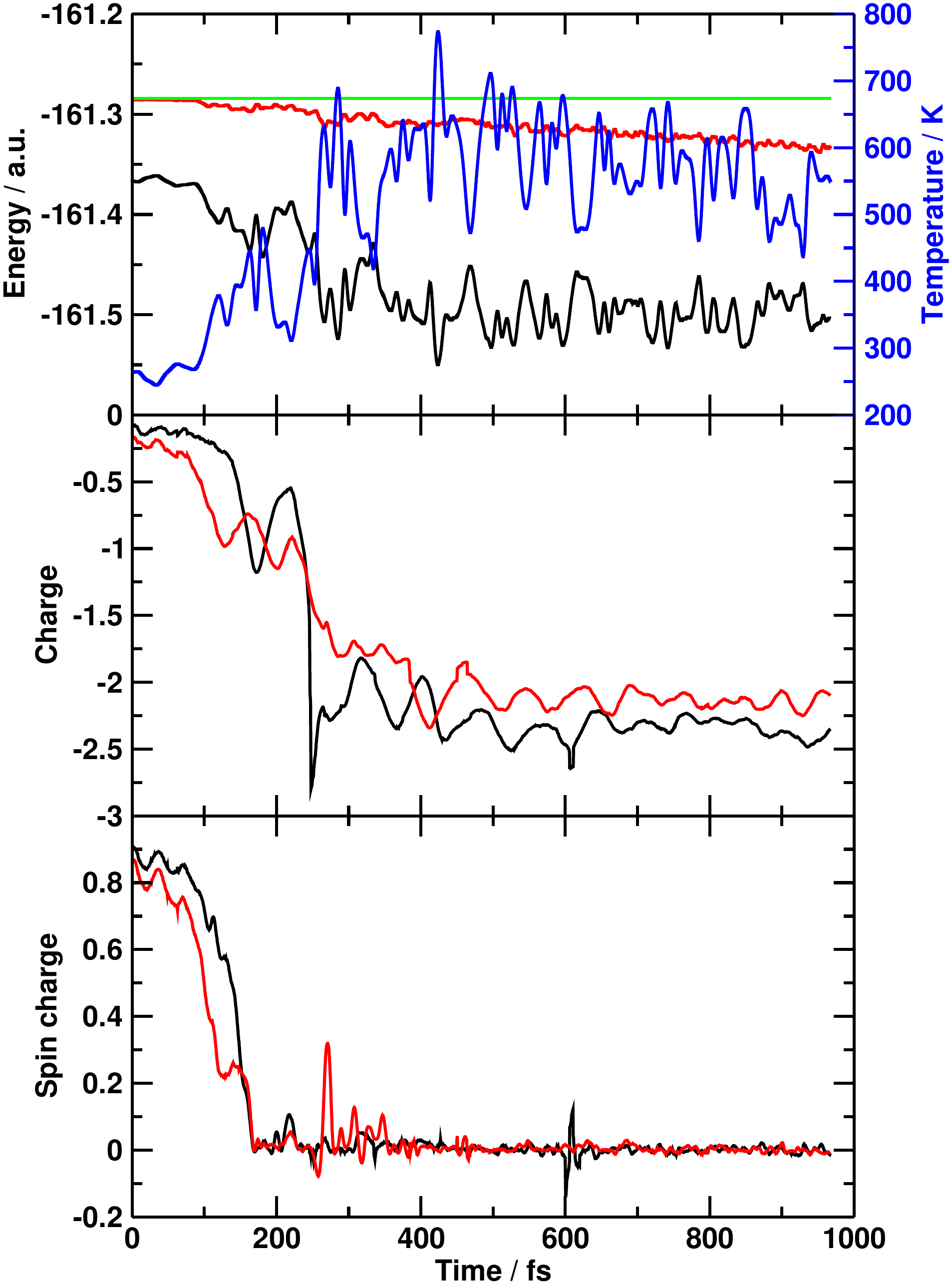}
  \caption{Snapshots from a simulation run showing a single oxygen molecule reacting
with the surface. The oxygen molecule (red) approaches the surface and dissociates.
The spin densities are shown in orange.
}
  \label{Fig3}
\end{figure}

\ref{Fig3} shows some snapshots of this simulation run. Before the
start of the reaction, the oxygen molecule is in its triplet ground state
which is reflected by a high spin density. The spin density is transferred to the
surface while the molecule binds to the surface and forms a three-membered ring
with an aluminium atom. (The total spin of the system stays 1). After the
spin charge of the oxygen atoms has decayed to zero, they dissociate (fourth snapshot
in \ref{Fig3}). Some 30 fs later there is again a certain accumulation of
spin density at the oxygen atoms which is obvious also from the peak in the
graph of the spin charge (\ref{Fig2}) about 270 fs after the end of the equilibration.
After this oscillation the oxygen atoms relax in surface lattice sites whereby an aluminium
atom is strongly disturbed and dislocated from its lattice site to another one (\ref{Fig3}, last snapshot).
From following the behaviour of the spin charge, it is obvious that the surface can
easily swallow the spin. The two unpaired spins avoid each other within the electron gas of the metal
resulting in a vanishing exchange interaction.
This finding is trivial: There is nothing like an aluminium triplet state which could energetically be
discriminated from the singlet state.

\begin{figure}
  \includegraphics[width=8cm]{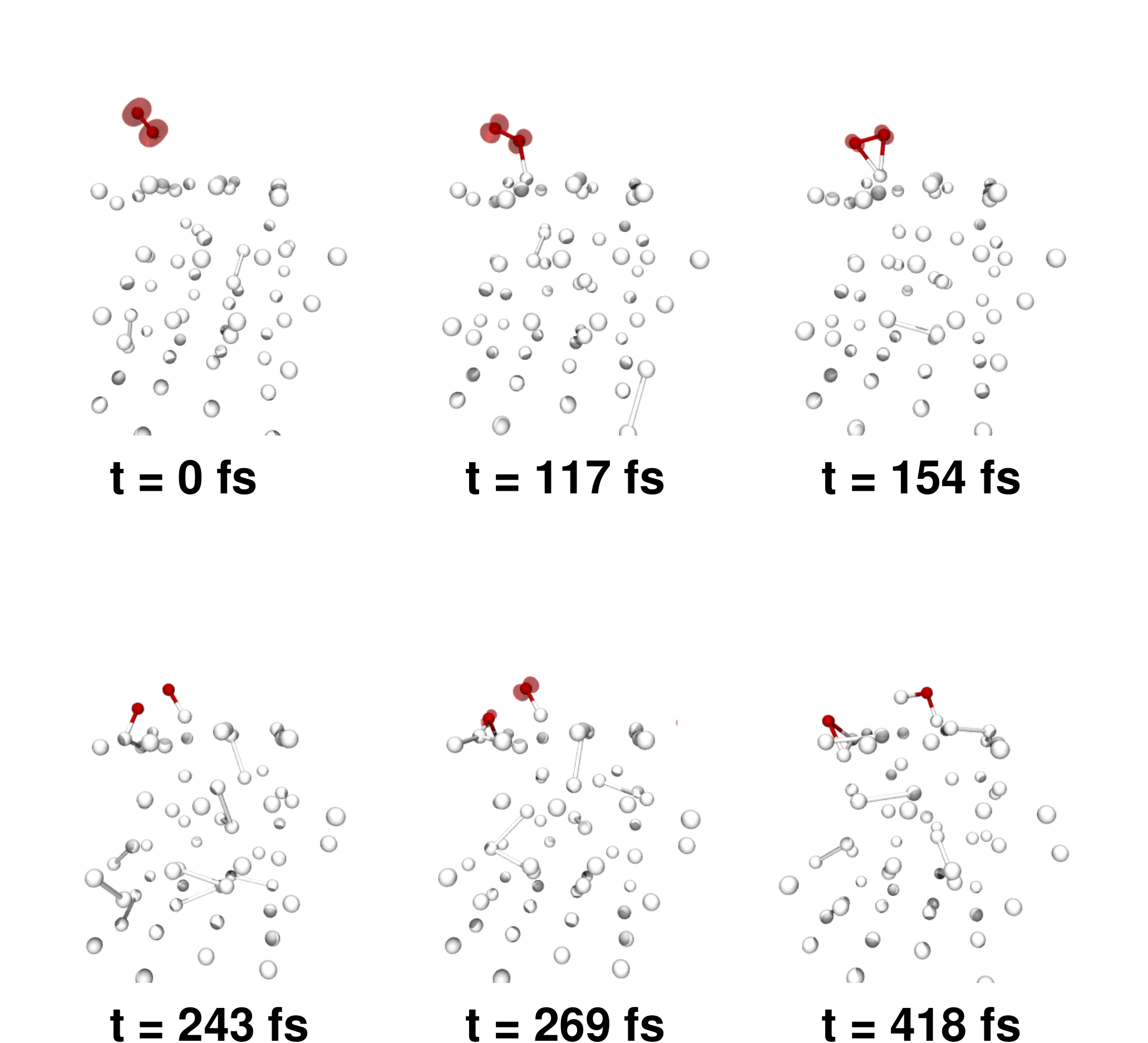}
  \caption{Change of the distance of the two oxygen atoms in the ten simulations and average value.
The average distance after dissociation and relaxation (0.4 nm) agrees well with the experimental value (0.5 nm).
}
  \label{Fig4}
\end{figure}

\ref{Fig4} shows the variance of the O-O distances during the reaction.
The motion of the oxygen atoms ends up at distances of about 
1 $a_s$, $\sqrt{3}$ $a_s$, 2 $a_s$, and $\sqrt{7}$ $a_s$ with $a_s$ = 2.86 \AA\, being the nearest neighbour spacing.
The distribution agrees excellently with experiment \cite{Schmid2001}. Note, however, that
the attribution of a certain distance to a certain
channel is not a hundred percent reliable as is illustrated
by the exemplary simulation run discussed above (\ref{Fig1}): Due to the strong disturbance
of the surface layer, a final distance of 5 \AA\, is reached, while from the distance of the relevant
lattice sites a distance of 7.6 \AA\, would be computed. 

The average distance obtained in the ten simulation runs is 0.4 nm.
This result is in nice agreement with the publication by Schmid et al.\ \cite{Schmid2001} who
report a mean interatomic distance after adsorption of 0.5 nm.
The deviation of the numerical value is within the error which
stems from the limitations in statistics,
simulation time and simulation cell size.
The computed value certainly does not agree, however, with the value of 8 nm reported in the work
by Brune et al.\, \cite{Brune1992,Brune1993}.
Due to the lower resolution of these early STM pictures adatom pairs were obviously interpreted as single atoms. \\
Experimentally, the distribution of the adatoms (or rather adatom pairs) was found to be random at low oxygen pressure while
at higher oxygen pressure the formation of islands was observed \cite{Brune1993}.
Also increasing the temperature supports this island formation \cite{Trost1997}.
From these experimental observations and from our results, island formation is less due to
the initial motion of the oxygen atoms till they are chemisorbed, than to the strong increase
of kinetic energy in the upper aluminium layer leading to partial melting of the metal.
In our picture it is not single oxygen atoms which move on the surface to form large islands, but
small and hard aluminium oxide islands which float on a soft metal surface.

\begin{figure}
  \includegraphics[width=8cm]{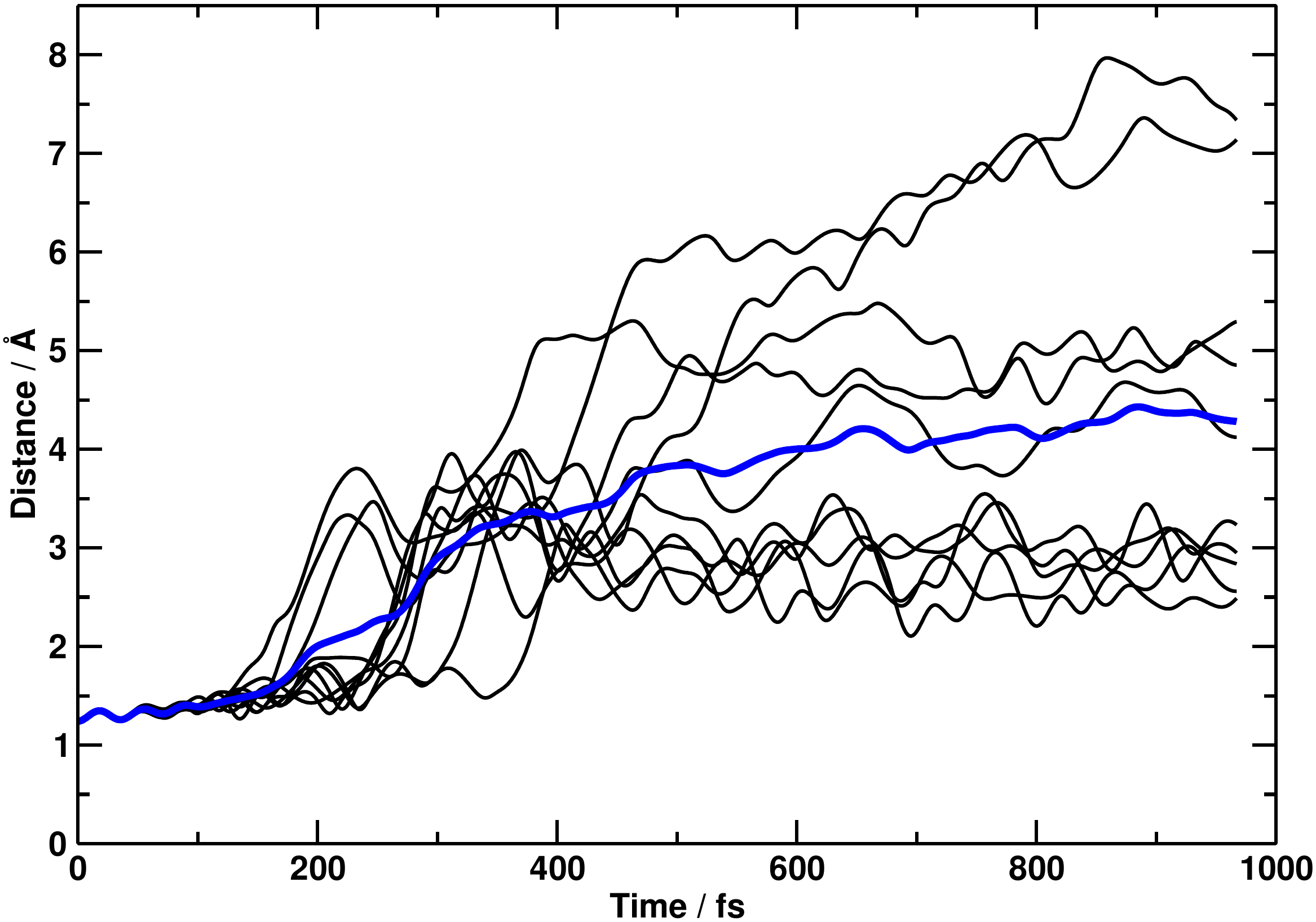}
  \caption{Snapshots from the molecular dynamics simulation
of the reaction of the aluminium surface with molecular oxygen.
}
  \label{Fig5}
\end{figure}

The beginning of a surface layer formation was studied in an additional simulation with liquid oxygen
simulated by 24 oxygen molecules inbetween the aluminium layers.
\ref{Fig5} shows some snapshots from a molecular dynamics simulation at 300
K. After the equilibration, the distance between the nearest
oxygen atoms and the surface is about 2.3 to 2.4 \AA. Nine out of 24 oxygen molecules in the simulation cell
react during the first
10000 steps (484 fs), four of them at the slab surface shown in the figure. 
A thermostat was used which
strongly reduces the kinetic energy set free in this extremely exothermic surface reaction.
Nevertheless, the strong disorder of the top-most aluminium layer
is obvious.
Statistically, similar numbers of oxygen atoms with two up spins and two down spins hit the surface,
so a large spin accumulation never happens.
Upon adsorption and dissociation, the oxygen atoms start to form a rigid surface layer
of triply coordinated adatoms illustrating the beginning of Al$_2$O$_3$ formation.

\section{Conclusions}

In conclusion, Car-Parrinello molecular dynamics simulations of the reaction
of molecular oxygen with aluminium show that
the oxygen molecules are chemisorbed immediately upon contact with the surface.
Dissociation leads to adatoms which are separated by 0.4 nm on average in very good
agreement with experiment \cite{Schmid2001}. 
The mechanism is best described as an adsorption-dissociation mechanism:
the oxygen atoms are chemisorbed, dissociate within a few 100 fs and relax within 
about a picosecond at lattice sites close-by. Mechanistically this is inbetween a dissociative
chemisorption and a 'hot adatom' mechanism. The resulting oxygen atoms are not 
necessarily located at neighbouring sites,
but, on the other hand, do not
move freely over the surface before relaxing.
Extensions of density functional theory are not necessary to explain
this reaction. The spin density of the triplet oxygen molecule
is simply transfered to the surface during the reaction.
Density functional theory within
the BLYP approximation turns out to be very well suited to describe this surface reaction.
The approach may serve to investigate many more experiments in this field.

\section{Methods}
For our molecular dynamics simulations we used the implementation of the
Car--Parrinello molecular dynamics (CPMD) scheme \cite{Car1985} in the CPMD
code \cite{cpmd}. This scheme uses density functional theory (DFT)
\cite{Hohenberg1964,Kohn1965} for the description of the electronic structure.
The unrestricted formulation of the Becke--Lee--Yang--Parr (BLYP)
\cite{Becke1988,Lee1988} generalized gradient approximation (GGA) functional
was used. In the plane wave code CPMD, pseudopotentials are used for the
description of the core electrons. The separable dual space Gaussian
pseudopotentials by Goedecker, Teter and Hutter (GTH)
\cite{Goedecker1996,Hartwigsen1998} were used with a plane--wave cutoff of 90
Rydberg. The fictitious electronic mass was set to the default value of 400
atomic units (a.u.). A small time step of 2 a.u. (0.048 fs) was chosen.
The temperature of the nuclei was set to 300 K.
The model system consists of four layers of 16 aluminium
atoms stacked in an ABCA order in an orthorhombic simulation cell describing a
(111)--surface of fcc aluminium. Previous test calculations showed that four layers
are sufficient to describe the chemistry of the system.
A lattice constant of $a_0 = 4.04959$ \AA\, ~was
used for calculating the cell parameters, corresponding to a fcc
nearest neighbour distance of $a_s = a_0 / \sqrt{2} = 2.86349 \AA$. 
Between the layers
a spacing of approximately 12.0 \AA\, is introduced.
The resulting cell parameters are
21.35213 \AA ~($4\frac{\sqrt{3}}{3}a_0+12.0 \AA$), 11.45396 \AA
~($4\frac{\sqrt{2}}{2}a_0$) ~and 9.91942 \AA ~($2\sqrt{\frac{3}{2}}a_0$). 
For the simulation of liquid oxygen, the spacing
between the layers is filled with 24 oxygen molecules corresponding to a density of
approximately 935  $\frac{kg}{m^3}$, which is  
a bit lower than that of liquid oxygen (1120 $\frac{kg}{m^3}$
\cite{Holleman1995}). As liquid oxygen is highly reactive and would react
immediately, it was replaced by unreactive
nitrogen molecules during the equilibration.
The temperature of the
reacting system would rise rapidly in the 
liquid oxygen simulation, hence
Nos\'{e}--Hoover thermostats
\cite{Nose1984,Nose1984a,Hoover1985} were used to control the temperature of
the nuclei as well as the fictitious kinetic energy of the electrons.
As thermostat parameters we use a frequency of 3000 cm$^{-1}$ for coupling the
nuclei to the bath and a frequency of 10000 cm$^{-1}$ for the electrons. The
fictitious kinetic energy of the electrons was chosen to be 0.07 a.u.
The charges and spin charges were calculated by integrating the densities
and spin densities, respectively, using
Bader analysis to determine the integration range.
\cite{Bader1990}.

\bibliography{../../literature.bib}

%
%
%

%

\end{document}